\documentclass[aps,prl,reprint,amsmath,amssymb,superscriptaddress]{revtex4-2}

\usepackage[british]{babel} 
\usepackage[dvipsnames, table]{xcolor}
\definecolor{mygray}{gray}{0.9}
\usepackage{caption}
\DeclareCaptionFormat{custom}{
\textbf{\footnotesize#1#2}\footnotesize#3
}
\usepackage{graphicx} 
\graphicspath{{Figures/}}

\usepackage{tabularx}
\usepackage{multirow}
\usepackage{colortbl}
\usepackage{array}  
\usepackage{esint}
\usepackage{amsmath}
\usepackage{amssymb}
\allowdisplaybreaks
\usepackage{mathtools} 
\usepackage{physics} 
\usepackage{BOONDOX-calo} 
\usepackage{bbm} 

\begin{document}

\title{Propagation of focused scalar and vector vortex beams in anisotropic media: a semi-analytical approach}

\author{Vittorio Aita}

\affiliation{Department of Physics and London Centre for Nanotechnology, King's College London, Strand, London WC2R 2LS, UK}

\author{Mykyta Shevchenko}

\affiliation{Department of Electronic \& Electrical Engineering, University College London (UCL), Torrington Place, London WC1E 7JE, UK}

\author{Francisco~J.~Rodr\'iguez-Fortu\~no}

\affiliation{Department of Physics and London Centre for Nanotechnology, King's College London, Strand, London WC2R 2LS, UK}

\author{Anatoly~V.~Zayats}

\affiliation{Department of Physics and London Centre for Nanotechnology, King's College London, Strand, London WC2R 2LS, UK}

\begin{abstract}
In the field of structured light, the study of optical vortices and their vectorial extension--vectorial vortex beams--has garnered substantial interest due to their unique phase and polarisation properties, which make them appealing for many potential applications. Combining the advantages of vortex beams and anisotropic materials, new possibilities for electromagnetic field tailoring and manipulation can be achieved in nonlinear optics, quantum and topological photonics. These applications call for a comprehensive modelling framework that accounts for properties of both anisotropic materials and vector vortex beams. In this paper, we describe a semi-analytical model that extends the vectorial diffraction theory to the case of focused vortex beams propagating through a uniaxial slab, considering both the cases of scalar and vectorial vortexes in the common framework of a Laguerre-Gaussian mode basis. The model aims to provide a comprehensive description of the methodology, enabling the implementation of complex beams transmission through, reflection from and propagation in uniaxial anisotropic materials for specific applications. As a demonstration of its versatility, we apply the developed approach to describe propagation of high-order vortex beams in uniaxial materials with various dispersion characteristics, exploring the elliptic, hyperbolic and epsilon-near-zero regimes. 
We show how variations of the medium anisotropy modify the beam structure due to the vectorial nature of their interaction, which results from the different permittivities of the medium for transverse and longitudinal field components. The applicability of the approach can be extended to artificially structured media if they can described by effective medium parameters. The developed formalism will be useful for modelling of interaction of complex beams with uniaxial materials, allowing a common framework for a large variety of situations, which can also be extended beyond the electromagnetic waves.

\end{abstract}

\date{\today}

\maketitle

\section{Introduction}
Since their first introduction~\cite{allen1992}, optical vortex beams (OVBs) have been subject of countless investigations and have led to important advancements, including in-cavity generation of OVBs, all-optical encryption techniques, OAM-based particles manipulation and beam shaping devices ~\cite{shen2019optical,naidoo2016,tang2018theoretical,wang2023coloured,piccardo2023broadband,afanasev2023nondiffractive}. The feature that makes optical vortices so interesting is their phase singularity, which is quantified by their topological charge ($\ell$): an integer describing how many times the phase wraps in $[0,2\pi]$ interval in a closed loop around the beam centre. A nonzero topological charge makes the wavefront of OVBs helicoidal, with the number of helices per wavelength distance determined by the value of $\ell$ and the handedness by its sign. Therefore, OVBs possess an orbital angular momentum (OAM) of $\pm\hbar\ell$ per photon. OVBs are important in optical communications, quantum optics and imaging, as they provide an additional degree of freedom for photons, introducing the possibility to encode more information in the same beam using topological charge. The combination of co-propagating OVBs leads to a non-uniform polarisation patterns in the resulting beam. In such, the so-called vectorial vortex beams (VVBs), the phase singularities of OVBs translate in polarisation singularities~\cite{maurer2007tailoring,zhan2009cylindrical}. These singularities are in many cases accompanied by strong longitudinal fields and often result in unusual behaviour, e.g., the violation of the optical theorem for scattering~\cite{krasavin-lsa} or formation of topological structures of light, such as optical skyrmions~\cite{shen2023topological}. 

As interesting as they are in a context of propagation in free space or uniform media, when interacting with anisotropic materials the behaviour of OVBs and VVBs becomes even more complex, leading to potential new opportunities and applications with new frontiers of beam shaping and polarisation control. The task of characterising their propagation in anisotropic media is far from trivial due to the inherent complexities arising from an interplay of the vectorial nature and vortex structure of the beams and the anisotropic properties of the material. While purely numerical simulations may provide the required information, the restrictions on the use of periodic boundary conditions due to the final size of the beam, results in the significant demands on computational resources and convergence issues. Therefore, the development of accurate and efficient modelling becomes imperative for studying and controlling the behaviour of VVBs in such scenarios, and advancing their applications.

In this work, we develop a semi-analytical approach which can be applied to the propagation of focused scalar and vectorial vortex beams through an anisotropic slab. The approach provides an extension of already examined cases of tightly focused beams propagating in free space and isotropic materials~\cite{mansuripur1986distribution,quabis2000focusing,youngworth2000focusing,biss2004cylindrical}. Both OVBs and VVBs are considered, with the former modelled as Laguerre-Gaussian beams of general order $\mathrm{LG}_{\ell p}$ and the latter as a superposition of orthogonally polarised OVBs with opposite topological charges $\pm\ell$~\cite{milione2011higher}. Following a brief introduction to the vector diffraction theory in multi-layered isotropic media, a detailed description of the methodology is presented for an anisotropic uniaxial medium. Examples of applications of the approach are given, exploring the cases of Laguerre-Gauss beams of different orders, propagating through various categories of uniaxial media. The developed approach will be a useful tool for modelling the interaction of complex beams with material systems of chosen optical properties. The cases covered by the developed method include both isotropic and anisotropic (limited to uniaxial) materials and allow the exploration of various dispersion regimes, elliptic, hyperbolic, as well as epsilon-near-zero.

\section{Vectorial diffraction theory in isotropic multi-layered media}
\begin{figure}[!t]
    \centering
    \includegraphics[width = 0.45\textwidth]{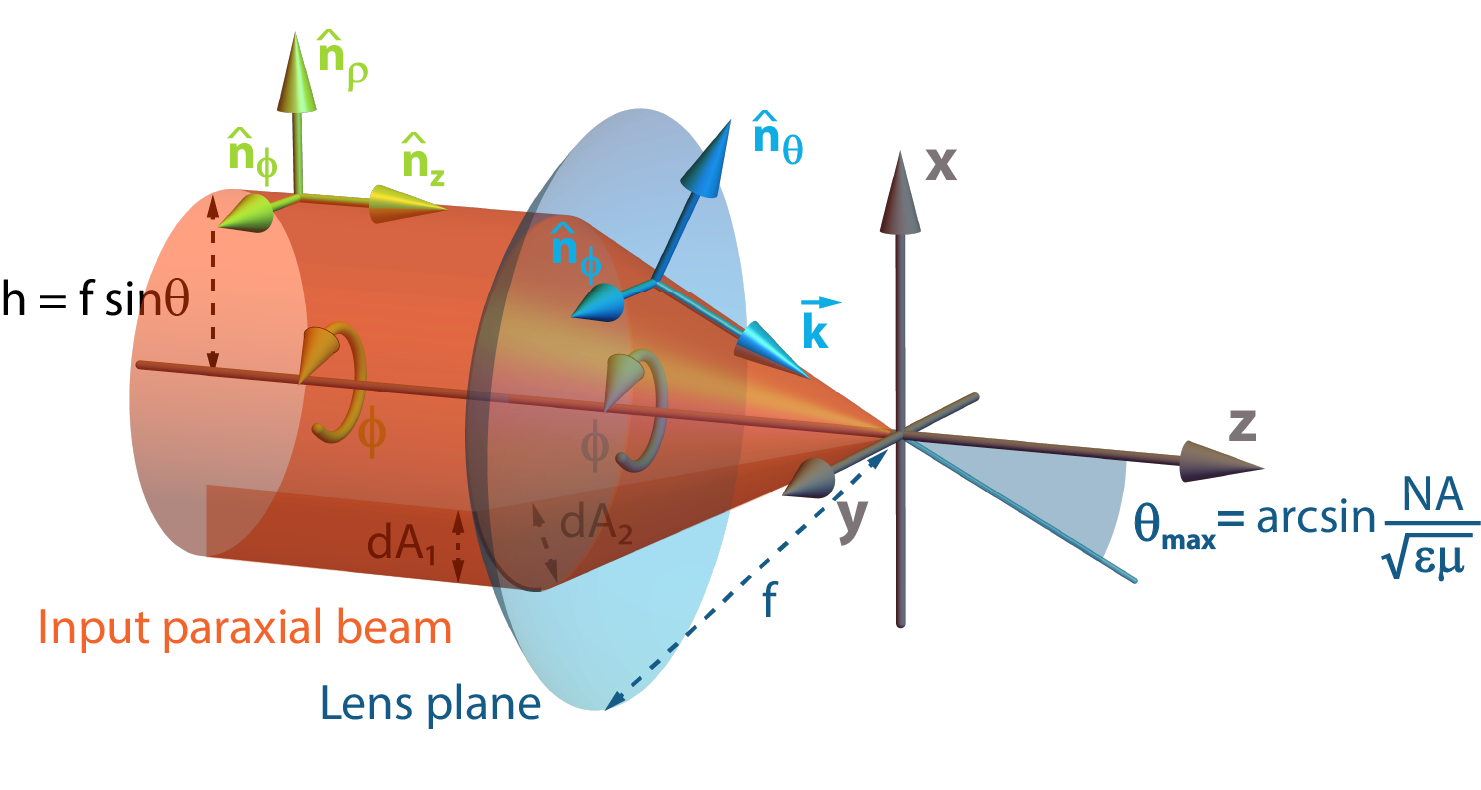}
    \caption{Schematics of a paraxial beam focused by a lens. The incoming beam, considered in cylindrical coordinates ($\rho,\phi,z$, shown in green), is mapped onto a spherical co-ordinate frame ($k,\phi,\theta$, shown in blue) upon refraction through a lens. Each paraxial ray at a height $h=f\sin\theta$ from the optical axis corresponds to a refracted ray propagating along the direction $\theta$. The angle $\theta$ is limited by the numerical aperture of the system to $\theta_{\rm max}$. 
    }
    \label{fig:rw_frame}
\end{figure}
The physical problem behind this approach is set to find the electric field of a focused beam, in a given volume divided into three domains, with the central one showing an anisotropic dielectric permittivity. One choice could be to follow a very general approach originally developed in~\cite{stratton1939diffraction} in the framework of the vectorial diffraction theory. It consists of an integral definition of the electric field at a fixed observation point, depending on the boundary electric and magnetic fields at the surface of an arbitrarily shaped aperture. For the majority of optical systems, the general approach can be simplified considering the asymptotic condition of the far-field diffraction, which can physically be thought of as the field distribution located sufficiently far from the focusing element of the system. Under this approximation, the Stratton-Chu integral can be understood as the electric field at a point $\va{r}$ in space originating from the superposition of an infinite number of plane waves propagating from the aperture to the point $\va{r}$. If the aperture is considered to be a spherical pupil, this approach can be reduced to the well-known Debye-Wolf integral~\cite{wolf1959electromagnetic,wolf1981conditions,sheppard1977}:
\begin{equation}\label{eq:DebWolf}
\va{E} (\va{r}) = - \frac{ i k \, e^{-i k f}}{2 \pi f} \iint \limits_{\Omega} \va{E}_{\Omega} \, e^{i \, \va{k}\cdot\va{r}} \,\dd\Omega\,,
\end{equation}
where $f$ is the distance between the point $\va{r}$ and the pupil, $\va{k} = k_0\va{k}_r = \omega/c \va{k}_r$, with $\va{k}_r$ being the ``relative'' wavevector in the medium, $\va{E}_{\Omega}$ is the electric field strength within the solid angle $\Omega$ under which the pupil is seen from $\va{r}$. The integration over $\Omega$ corresponds to a summation of all the plane waves directed to the point  $\va{r}$. Considering a decomposition of the unknown field into a superposition of plane waves with propagation properties dictated by the structure of the system, the angular spectrum formalism can be developed, which is the basis of the Richards-Wolf (RW) theory of vectorial diffraction~\cite{RichardsWolf}. 

A general solution of the wave equation can be written as the superposition of a number of plane waves with varying wavevector $\va{k}$, weighted by an appropriate $\va{A}$ function~\cite{whittaker1902course,wolf1959electromagnetic, novotnyChap2,mandel_wolf_1995}:
\begin{subequations}\label{eq:ang_spec}
\begin{align}
    \va{E}(x,y,z) &=  \iint\limits_{-\infty}^\infty \va{A}(k_x,k_y)\,e^{i(k_xx + k_yy)}\,e^{\pm i k_z z}\dd k_x\,\dd k_y\,,\\
    \va{A}(k_x,k_y) &= \frac{1}{4\pi^2}\iint\limits_{-\infty}^\infty \va{E}(x,y,0)\,e^{i(k_xx + k_yy)}\dd x\,\dd y\,,
\end{align}    
\end{subequations}
\noindent where the function $\va{A}(k_x,k_y;z)$, which is the angular spectrum of $\va{E}(\va{r})$, is the weight of each plane wave component propagating at the direction $\va{k}$. It corresponds to the two-dimensional Fourier transform of the field, calculated at a reference plane (here $z = 0$). This plane can in principle be a plane orthogonal to an arbitrary direction, but it is usually conveniently chosen to be the wave propagation direction. For uniform and isotropic medium, considering a time-harmonic field, the propagation in real space can be calculated as a product in reciprocal space:
\begin{gather}
    \va{A}(k_x,k_y;z) = \va{A}\!\left(k_x,k_y,0\right)e^{\pm ik_z z}\,, \\
    \quad k_z\equiv k_0\sqrt{1-k_{\mathrm{r}x}^2-k_{\mathrm{r}y}^2}\,.
\end{gather}
This represents a powerful formalism for the computation of an electromagnetic field, assuming its distribution can be written in a reference plane. It can also be easily combined with the description of multilayered media by including the Fresnel coefficients describing the system in Eq.~(\ref{eq:ang_spec}). For the simple case of multilayered media where all the interfaces are planes parallel to each other and orthogonal to the propagation direction, the boundary conditions can be solved to find the Fresnel coefficients. In the scope of this paper, the discussion is limited to a slab with three consecutive materials of uniform dielectric media, but the formalism is not restricted to this simplification. Within this assumption, the Fresnel coefficients are (see, e.g.,~\cite{lekner1994optical}): 
\begin{subequations}\label{eq:fcs_2int}
        \begin{align}
           r^{s,p}_\mathrm{I}  &=\frac{e^{2 i k_{1z} z_0}\  \left(r_{12}^{s,p}+r_{23}^{s,p} e^{2 i d k_{2z}}\right)}{1+r_{12}^{s,p} r_{23}^{s,p} e^{2 i d k_{2z}}}\,,\\
             t^{s,p}_\mathrm{I}  &=\frac{e^{i(d(k_{2z} - k_{3z})+z_0(k_{1z} - k_{3z}))}\ t_{12}^{s,p}t_{23}^{s,p}}{1+r_{12}^{s,p} r_{23}^{s,p} e^{2 i d k_{2z}}} \,,\\
            r^{s,p}_\mathrm{II}  &=\frac{e^{i (2 d k_{2z}+z_0 (k_{1z}+k_{2z}))}\ r_{23}^{s,p}t_{12}^{s,p} }{1+r_{12}^{s,p} r_{23}^{s,p} e^{2 i d k_{2z}}} \,,\\
             t^{s,p}_\mathrm{II}  &=\frac{e^{i (d (k_{2z}-k_{3z})+ z_0 (k_{1z}-k_{3z}))}t_{12}^{s,p} t_{23}^{s,p} }{1+r_{12}^{s,p} r_{23}^{s,p} e^{2 i d k_{2z}}} \,,
    \end{align}
\end{subequations}
where the numeric subscript labels the medium (two number pairs $ij$ refer to the interface between medium $i$ and $j$), so that $r_{ij}$ and $t_{ij}$ are the Fresnel reflection and transmission coefficients, respectively, for the interface
 between the media $i$-th and $j$-th, subscripts I and II refer to the slab interfaces located at $z=z_0$ and $z=z_1$, respectively, $d = |z_0-z_1|$ is the thickness of the slab, and $s$ and $p$ are the polarisation components of the field with respect to the plane of incidence and the direction of $\va{k}$. The electric field in the integration volume is piece-wise defined as
\begin{equation}\label{eq:piecewise_field}
    \va{E}(\va{r}) = \begin{dcases}
        \va{E}_1 = \va{E}_\mathrm{i} + \va{E}_{\mathrm{r}1} & \mathrm{for}\quad  z \leq z_0\\
        \va{E}_2 = \va{E}_{\mathrm{t}1} + \va{E}_{\mathrm{r}2}& \mathrm{for}\quad z_0< z \leq z_1\\
        \va{E}_3 = \va{E}_{\mathrm{t}2}& \mathrm{for}\quad z>z_1
    \end{dcases}\,,
\end{equation}
where the labels i, r, t stand for ``incident'', ``reflected'' and ``transmitted'', respectively. Each term is then projected onto its $s$ and $p$ field components, relating, according to Eq.~(\ref{eq:ang_spec}), its angular spectrum to the incident field via the Fresnel coefficients from Eqs.(\ref{eq:fcs_2int}). The field inside the slab is then 
\begin{subequations}\label{eq:rw_slab_fcs}
    \begin{gather}
        \va{E}_2 = \va{E}_{\mathrm{t}1} + \va{E}_{\mathrm{r}2} = \va{E}_{\mathrm{t}1}^s + \va{E}_{\mathrm{t}1}^p +  \va{E}_{\mathrm{r}2}^s + \va{E}_{\mathrm{r}2}^p\,,\\
        \va{A}_{\mathrm{t}1}^{(s,p)} = t^{s,p}_\mathrm{I}\,  \va{A}_\mathrm{I}^{(s,p)}\,,\qquad \va{A}_{\mathrm{r}2}^{(s,p)} = r^{s,p}_\mathrm{II}\,  \va{A}_\mathrm{I}^{(s,p)}\,.
    \end{gather}
\end{subequations}

A useful application of the angular spectrum formalism is the study of focused fields, which has led to the RW theory~\cite{wolf1959,RichardsWolf}. In simple terms, the field of a focused laser beam is determined by the effect of the focusing system--a lens--on the incoming beam, which is described by a boundary problem at the interface corresponding to the lens. 
The focusing system is considered to be aplanatic and light comes from a source at infinity, so that the wavefront is assumed to be planar in the plane of the lens. An implicit approximation of this method comes from the use of the angular spectrum formalism in the asymptotic regime: the reference plane is located at infinity, where the wavefront is planar. Assuming the beam is paraxial before the lens, the reference plane can be chosen to coincide with the lens surface. Within asymptotic approximation, the fields in the proximity of the lens can then be formulated in the frame of geometrical optics with two conditions to be fulfilled~\cite{novotnyChap3}: (i) the sine condition and (ii) the intensity law. The latter is related to conservation of energy upon propagation through the lens: the energy flux of each ray needs to be constant and the power has to be the same on both side of the lens surface. The former describes the lens boundary as a sphere centred in its geometric focus, with the focal length being its radius, and ensures a one-to-one mapping of the rays incident on the lens to corresponding refracted rays. The distance of each ray from the optical axis--chosen parallel to the propagation direction--can be written as $h = f\sin\theta$ (Fig.~\ref{fig:rw_frame}) so that independently of the choice of coordinates in the half-space before the lens, it can be mapped onto spherical coordinates on the sphere surface. The angle $\theta$ is the refraction angle of the ray at the distance $h$ from the optical axis. 

The above approximations lead to a modification of Eq.~\eqref{eq:DebWolf}: 
\begin{subequations}\label{eq:focus_field}
\begin{align}\nonumber
    &\va{E}(\rho,\varphi,z) = -\frac{ikfe^{-ikf}}{2\pi} \\&\times\int\limits_{0}^{\theta_{\mathrm{max}}}\int\limits_{0}^{2\pi}\va{E}_\infty(\theta,\phi) \, \underbrace{e^{ikz \cos\theta}e^{ik\rho\sin\theta\cos(\phi-\varphi)}}_{e^{i\va{k}\cdot\va{r}}}\overbrace{\sin\theta \,\dd\phi\,\dd\theta}^{\dd\Omega} \,,\\
    \va{E}_\infty &= \Bigl[t^s\bigl(\va{E}_\mathrm{L}\vdot\vu{n}_\varphi\bigr)\vu{n}_\varphi + t^p\bigl(\va{E}_\mathrm{L}\vdot\vu{n}_\rho\bigr)\vu{n}_\theta\Bigr]\sqrt{\frac{n_1}{n_2}}\sqrt{\cos\theta}\,,
\end{align}
\end{subequations}
where $\va{E}_\mathrm{L}$ is the electric field incident on the surface of the lens. 
The integration in the reciprocal space has been limited to the angle $\theta_{\mathrm{max}}$ corresponding to the half aperture of the lens field of view ($\mathrm{NA} = n\sin\theta_{\mathrm{max}}$), while $\phi$ assumes all the possible values in $[0,2\pi]$ interval. The solution propagating along negative $z$ has been discarded for obvious physical reasons. The choice of spatial coordinates $(x = \rho\cos\varphi,\,y=\rho\sin\varphi)$ is made for simplifying the integrals computation. Because of the last exponential factor, it is in fact possible to use the integral definitions of modified Bessel functions of the first kind given by~\cite{RichardsWolf}
\begin{equation}\label{eq:besseltrig}
    \int\limits_0^{2\pi} \!\! \mqty(\cos(m\phi)\\\sin(m\phi))e^{ix\cos(\phi-\varphi)}\,\dd\phi = 2\pi i^{m}\,J_{m}(x)\mqty(\cos(m\varphi)\\\sin(m\varphi)),
\end{equation}
which allows for an analytical solution for at least one of the integrals, once the amplitude of the $\va{E}_\infty$ components are written as a combination of trigonometric functions. In order to solve Eq.~\eqref{eq:focus_field}, the $\va{E}_\infty$ field distribution needs to be made explicit. Focusing through \textit{isotropic} multilayered media can easily be implemented, following the procedure described in \cite{quabis2000focusing,youngworth2000focusing}.

\section{Vector diffraction theory in uniaxial media}
\begin{figure}[!t]
    \centering
    \includegraphics[width = 0.45\textwidth]{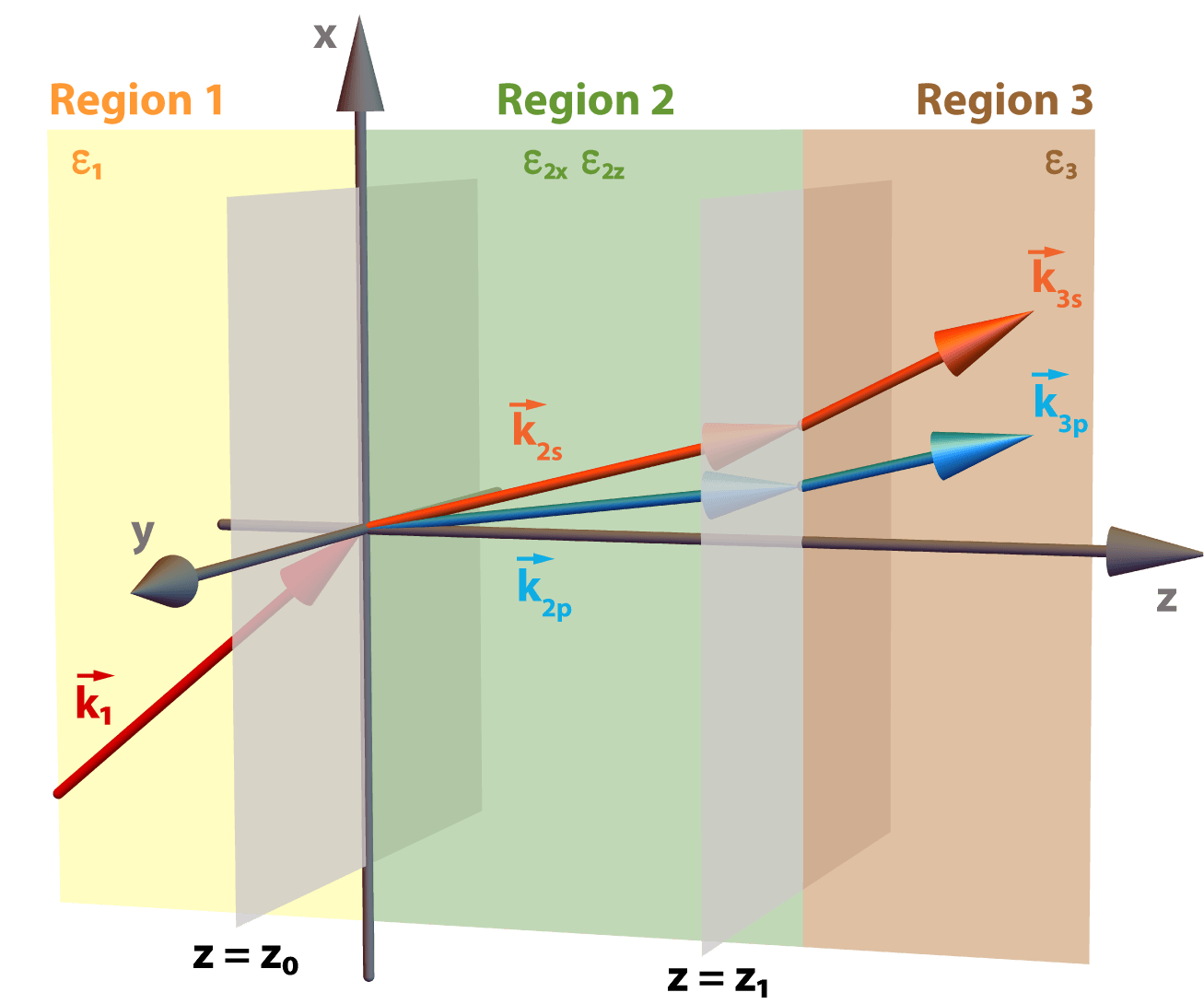}
    \caption{Geometry of the three-layered system. Light is incident from medium $1$, region $2$ is filled with a uniaxial material.}
    \label{fig:aniso_slab}
\end{figure}
In solving the boundary problem for an isotropic multi-layered medium, the conservation of the transverse component of the wave vector $(k_x,k_y)$ must be ensured, and its $z$ component can be represented through $k_x$ and $k_y$, once the dispersion relation of each medium is known. If the medium is anisotropic, this is generally no longer the case. For this reason, we limit our discussion to the case of a uniaxial crystal with the optical axis along the propagation direction ($\vu{z}$), so that the electric field in the $(x,y)$-plane sees an isotropic medium. In this instance, the permittivity tensor which describes the slab is given by
\begin{equation}\label{eq:epsilon}
    \boldsymbol{\varepsilon}_2 = \mqty(\dmat[0]{\varepsilon_{x},\varepsilon_{x},\varepsilon_{z}})\,.
\end{equation}
Each plane wave component in the angular spectrum has to satisfy the wave equation $\va{k}\cross\va{k}\cross\va{E}+\mu\varepsilon\va{E}=0$ which, using Eq.~\eqref{eq:epsilon} and reducing the system to a two-dimensional problem ($k_{\mathrm{r}y}=0$) can be written as:
\begin{equation}\label{eq:wave_aniso}
 \mqty[k_{\mathrm{r}z}^2 - \mu\varepsilon_x & 0 & -k_{\mathrm{r}x}k_{\mathrm{r}z}\\0 &(k_{\mathrm{r}x}^2 + k_{\mathrm{r}z}^2)- \mu\varepsilon_x & 0\\ -k_{\mathrm{r}x}k_{\mathrm{r}z}& 0 & k_{\mathrm{r}x}^2 - \mu\varepsilon_z]\va{E} = 0\ ,
\end{equation}
where the subscript $r$ denotes the ``relative'' wavevector ($\va{k} = k_0\va{k}_r = \left(\omega/c\right)\,\va{k}_r$), as previously mentioned.  The solutions for the electric field $\va{E}$ are then given by the null space of the $k$-matrix, which only exists when its determinant is zero. This condition can be fulfilled in two cases, corresponding to two possible polarisation modes. The associated magnetic field can be obtained from the Faraday's law ($\va{H} = (1/\omega \mu_0 \mu) \va{k} \cp \va{E}$). These two modes and their fields are presented in Table~\ref{tab:modes}.
\begin{table*}
\setlength\extrarowheight{4pt}
\captionof{table}{Modes of the electromagnetic field in a uniaxial material.}
\centering
    \begin{tabular}[t]{cccc}
    \arrayrulecolor{black}\hline
   \rowcolor{mygray}
   \textbf{Polarisation} & \textbf{Condition} & \textbf{E-field} & \textbf{H-field}\\[3pt]\hline
    \textit{s}& $ k_{\mathrm{r}x}^2 + k_{\mathrm{r}z}^2 = \mu\varepsilon_x $ & $\va{E}^s = \mathrm{E}_0^s\mqty(0\\1\\0)$& $\va{H}^s = \frac{\mathrm{E}_0^s}{c\mu_0\mu}\mqty(\mp k_{\mathrm{r}z}\\0\\k_{\mathrm{r}x})$\\[3pt]
     \arrayrulecolor{mygray}\hline
    \textit{p}&$\frac{k_{\mathrm{r}x}^2}{\varepsilon_z} + \frac{k_{\mathrm{r}z}^2}{\varepsilon_x} = \mu$ & $\va{E}^p = \frac{\mathrm{E}_0^p}{\sqrt{\mu\varepsilon_x}}\mqty(\pm k_{\mathrm{r}z}\\0\\\mp \frac{\varepsilon_x}{\varepsilon_z}k_{\mathrm{r}x})$& $\va{H}^p = \frac{\varepsilon_x \mathrm{E}_0^p}{\omega\mu_0\sqrt{\mu\varepsilon_x}}\mqty(0\\1\\0)$\\[3pt]
 \arrayrulecolor{black}\hline
    \end{tabular}
\label{tab:modes}
\end{table*}

The solutions in Table~\ref{tab:modes} suggest the possibility to define a basis with respect to the $k$-vector onto which the electric and magnetic fields~\cite{picardi2017unidirectional,Jackson} may be decomposed:
\begin{subequations}\label{eq:uniax_basis}
    \begin{gather}
         \vu{e}_s^\pm=\frac{1}{\sqrt{\mu\varepsilon_x}}\Bigl(\mp k_{\mathrm{r}y},\pm k_{\mathrm{r}x},0\Bigr)\,,\\
         \vu{e}_p^\pm = \frac{1}{\mu}\left(\pm \frac{k_{\mathrm{r}x}k_{\mathrm{r}z}}{\varepsilon_x},\pm\frac{ k_{\mathrm{r}y}k_{\mathrm{r}z}}{\varepsilon_x},\mp\frac{ k_\mathrm{rT}^2}{\varepsilon_z}\right)\,.
    \end{gather}
\end{subequations}
The core difference between the iso-~\cite{picardi2017unidirectional} and anisotropic cases stems from the different solutions found for the $k$-vector: upon propagation from Region 1 to Region 2 (Fig.~\ref{fig:aniso_slab}), an isotropic slab will make all the components of $\va{k}$ scale equally. In contrast, there is mixing of transverse and longitudinal components of $\va{k}$ for an anisotropic slab. The decomposition of the electric field into its $s$ and $p$ components allows for an easy implementation of the angular spectrum formalism for an uniaxial slab. Knowing the expressions for $k_z$ from the modes found for a uniaxial dielectric, the Fresnel coefficients in Eqs.~\eqref{eq:fcs_2int} can be adapted to an anisotropic layer.

\section{Modelling approach}

Equation \ref{eq:uniax_basis} can be applied to an anisotropic slab following the same procedure as in the isotropic case described above, with the Fresnel coefficients of the system included in the integrals in Eq.~\eqref{eq:focus_field}.
Each field appearing in the piece-wise definition in Eq.~\eqref{eq:piecewise_field} can be calculated once the explicit form of $\va{E}_\infty$ is fixed. We apply this semi-analytical approach to both the cases of scalar optical vortices and vector vortex beams. 

Both these type of beams can be described as Laguerre-Gaussian beams with a complex amplitude LG$_{\ell,p}$~\cite{SalehTelch3}:
\begin{subequations}
\begin{align}
    |\mathrm{LG}_{\ell,p}| &= \frac{w_0}{w(z)}\sqrt{\frac{2 p!}{\pi(p+|\ell|)!}}\Biggl(\frac{2\rho^2}{w(z)^2}\Biggr)^{|\ell|/2}\nonumber\\
    &\times \mathbb{L}_{\ell p}\Biggl(\frac{2\rho^2}{w(z)^2}\Biggr)\,,\\
    \arg\bigl(\mathrm{LG}_{\ell,p}\bigr)&=i(2p+\ell+1)\arctan\biggl(\frac{z}{z_r}\biggr)-\frac{\rho^2}{w(z)^2}\nonumber\\
    &-i \, k\frac{\rho^2}{R(z)}-i\ell\phi\,,
    \end{align}
\end{subequations}
where $\mathbb{L}_{\ell p}$ represents the generalised Laguerre polynomials~\cite{abramowitz1968handbook}, with $\ell$ and $p$ being their azimuthal and radial orders, respectively, $w(z)$ describes the beam lateral size as a function of the $z$ coordinate with $w_0$ being its minimum value, $R(z)$ is the wavefront radius of curvature, and $z_R = \pi w_0^2/\lambda$ is the beam Rayleigh range.
Following the prescriptions of the angular spectrum formalism, the reference plane is chosen as $z=0$, so that $w(z)=w_0$, $\exp(\rho^2/R)\rightarrow1$, and $\arctan(z/z_r)$=0. Finally, to account for the limitation to the field of view of the focusing element imposed by its finite size, the apodisation function $f_w(\theta) = \exp[-\frac{1}{f_0^2}\frac{\sin[2](\theta)}{\sin[2](\theta_{\mathrm{max}})}]$ should be introduced. Here, $f_0 = \frac{w_0}{f\sin\theta_\mathrm{max}}$ is a geometrical factor representing the ratio between the beam and the lens lateral sizes. With this choice, the dependence on $\rho$ is also modified so that $\rho^2 = \frac{w_0\sin[2](\theta)}{f_0\sin[2](\theta_{\mathrm{max}})}$. With these approximations, the mode amplitude only depends on the variable $\theta$, while its phase only on $\phi$:
\begin{equation}
    \mathrm{LG}_{\ell p} = \mathrm{C}_{\ell p}f_w(\theta)\,\Theta(\theta)\,\Phi(\phi)\,,
\end{equation}
where $\Theta$ and $\Phi$ collect the terms depending on $\theta$ and $\phi$, respectively, and $\mathrm{C}_{\ell p}$ all the constant factors:
\begin{subequations}
    \begin{gather}
        \mathrm{C}_{\ell p} = \sqrt{\frac{2 p!}{\pi(p+|\ell|)!}}\\
        \Theta(\theta) = \Biggl(\frac{2 \sin[2](\theta)}{w_0 f_0\sin[2](\theta_{\mathrm{max}})}\Biggr)^{|\ell|/2}\mathbb{L}_{\ell p}\Biggl(\frac{2 \sin[2](\theta)}{w_0 f_0\sin[2](\theta_{\mathrm{max}})}\Biggr)\\
        \Phi(\phi) = \mqty(\Phi_x(\phi)\\\Phi_y(\phi))\,,
    \end{gather}
\end{subequations}
where the two-dimensional vector ($\Phi_x,\,\Phi_y$) is needed for the description of vectorial vortices. The electric field incident on the lens ($\va{E}_\mathrm{L}$) can then be written in a general way as: 
\begin{equation}\label{eq:Ei_gen}
    \va{E}_\mathrm{L} = \mathrm{C} f_w(\theta)\,\Theta(\theta)\,\mqty(\Phi_x(\phi)\\\Phi_y(\phi))\,,
\end{equation}
where the exact definition of ($\Phi_x,\,\Phi_y$) depends on the type of vortex.

\subsection{Scalar vortex} 

In the case of a scalar vortex, $\Phi_x = \mathrm{E}_{0x}\exp(-i\ell\phi)$, $\Phi_y = \mathrm{E}_{0y}\exp(-i\ell\phi)$, and the dependence on $\phi$ is the same for both components resulting in uniform SOP. The beam polarisation can then be completely described by the coefficients $(\mathrm{E}_{0x},\mathrm{E}_{0y})$, which represent the projections of the SOP on the polarisation basis $\{\ket{\mathrm{H}},\,\ket{\mathrm{V}}\}$. Denoting these projections as $\bra{\rm H}\ket{\psi_s}$ and $\bra{\rm V}\ket{\psi_s}$, the scalar SOP $\ket{\psi_s}$ can be obtained as 
\begin{equation}\label{eq:SVB_SOP}
    \ket{\psi_s} = \ket{\rm H}\bra{\rm H}\ket{\psi_s}+\ket{\rm V}\bra{\rm V}\ket{\psi_s}\,.
\end{equation}

\subsection{Vector vortex}

Vectorial vortexes can be modelled as a superposition of LG beams with opposite values of $\ell$, the same value of $p$, and orthogonal circular polarisations. As a consequence of this, the dependence on $\theta$ of both components of $\va{E}_\mathrm{L}$ is the same, while the phase terms have different dependence on $\phi$. After introducing the polarisation basis~\cite{milione2011higher}
 \begin{subequations}\label{eq:circ_basis}
 \begin{align}
        \ket{\rm R,\ell} &= \frac{\ket{\rm H} - i \ket{\rm V}}{\sqrt{2}}e^{i\ell\phi}\,,\\
        \ket{\rm L,\ell} &= \frac{\ket{\rm H} + i \ket{\rm V}}{\sqrt{2}}e^{-i\ell\phi}\,,
    \end{align}
\end{subequations}
the vectorial SOP $\ket{\psi_v}$ can be written as the sum of orthogonal circular scalar vortices:
\begin{equation}\label{eq:vec_sops}
    \ket{\psi_v} = \ket{\rm R,\ell}\bra{\rm R,\ell}\ket{\psi_v}+\ket{\rm L,\ell}\bra{\rm L,\ell}\ket{\psi_v}\,,
\end{equation}
where the projections on the helicity basis (Eq.~\ref{eq:circ_basis}) are $\mathrm{E}_{0R} = \bra{\rm R,\ell}\ket{\psi_v}$ and $\mathrm{E}_{0L} = \bra{\rm L,\ell}\ket{\psi_v}$. To consistently label SOPs in Eq.~\ref{eq:vec_sops} and Eq.~\ref{eq:SVB_SOP}, 
the coefficients $\mathrm{E}_{0R}$, and $\mathrm{E}_{0L}$ in $\ket{\psi_v}$ can be expressed in terms of $\mathrm{E}_{0x}$ and $\mathrm{E}_{0y}$, so that
\begin{equation}
    \ket{\psi_v} = \frac{\mathrm{E}_{0x}+i \mathrm{E}_{0y}}{\sqrt{2}}\ket{\mathrm{R},\ell} +\frac{\mathrm{E}_{0x}-i \mathrm{E}_{0y}}{\sqrt{2}}\ket{\mathrm{L},\ell}\,.
\end{equation}
Within this notation framework, assuming $\ell = 0$, both $\ket{\psi_s}$ and $\ket{\psi_v}$ return uniform SOPs, expressed with the two-dimensional Jones vector $(\mathrm{E}_{0x},\mathrm{E}_{0y})$. For example, $(\mathrm{E}_{0x},\mathrm{E}_{0y}) = (1,0)$ gives the SOP $\ket{\rm H}$ and $(1,-i)/\sqrt{2}$ corresponds to $\ket{\rm R}$. Values of $\ell\neq 0$ will otherwise produce scalar SOPs from $\ket{\psi_s}$ and vectorial ones from $\ket{\psi_v}$.

\begin{figure*} [!t]
    \centering
    \includegraphics[scale = 1]{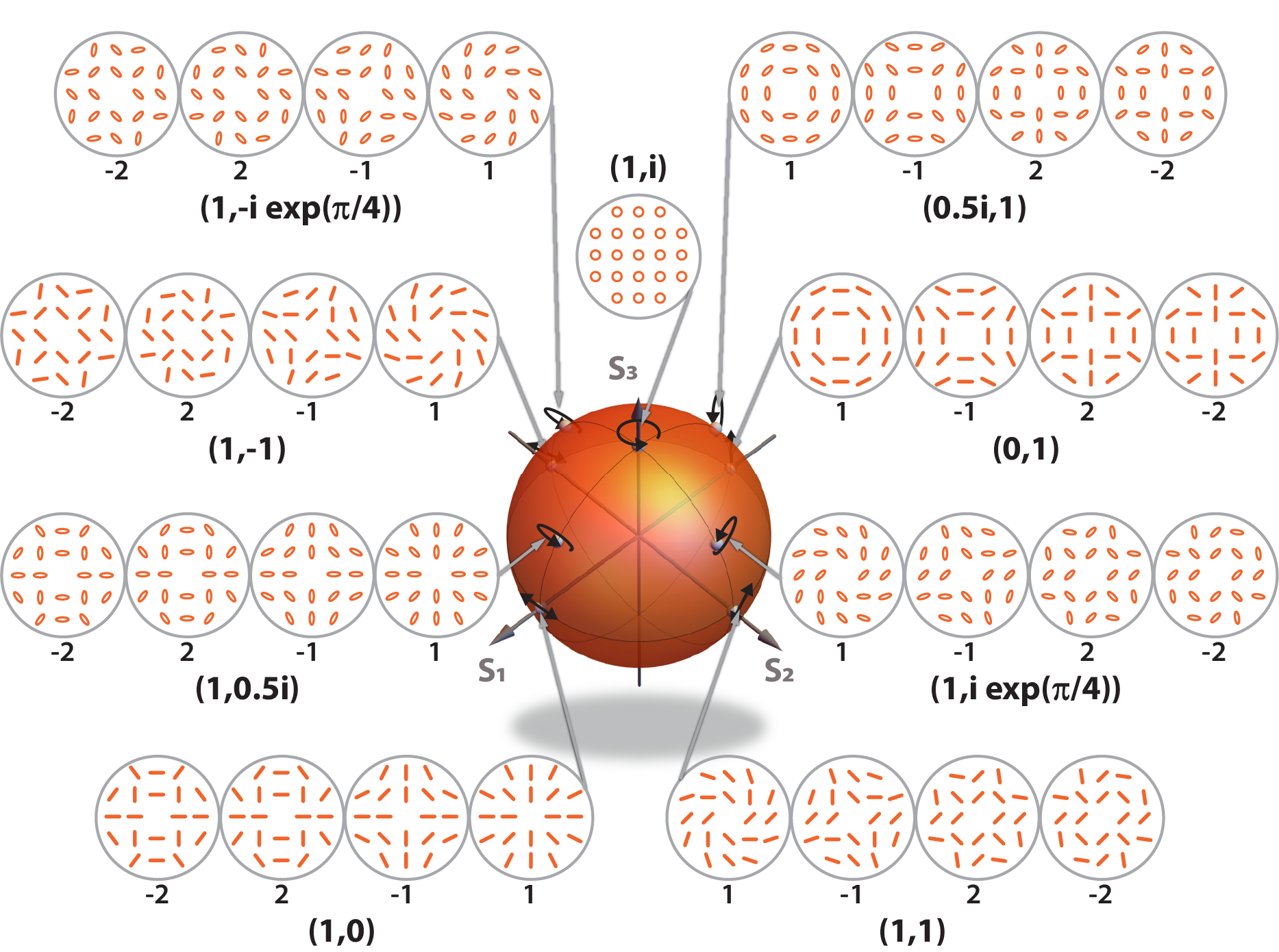}
    \caption{Poincaré sphere representation of the polarisation states. The Stokes vectors $S_1$, $S_2$ and $S_3$ are used as the coordinate axes of the polarisation space. Scalar polarisation states are represented on the surface of the sphere as black arrows, while vectorial SOPs, calculated according to Eq.~\eqref{eq:vec_sops} with values of topological charge $\pm$1, are shown in correspondence of the scalar SOPs having the same values of ($\mathrm{E}_{0x},\mathrm{E}_{0y}$). The number below each circle indicates the topological charge of the SOP above it, while the two-dimensional vectors indicate the value of $\mathrm{E}_{0x},\mathrm{E}_{0y}$ used for the corresponding series of SOPs. Similar SOPs can be obtained for the southern hemisphere of the Poincaré sphere with the only difference of a $\pi/2$ rotation of the pattern. North and south poles represent uniform circular polarisation regardless of the topological charge.}
    \label{fig:sops}
\end{figure*}

The examples of various SOPs are shown in Fig.~\ref{fig:sops}. The SOPs corresponding to $\ell=0$, which can be equivalently obtained as scalar vortices or ``degenerate'' vectorial ones, are presented as points on the surface of the sphere. Each point (E0x, E0y) on the Poincaré sphere can also represent a vectorial polarisation state, if a value of $\ell\neq 0$ is chosen. For example, the point associated with $\ket{H,0}$, which in a scalar case corresponds to a horizontal SOP, represents instead a radial beam $\ket{H,1}$ for $\ell$ = 1. SOPs belonging to the same series show the same local polarisation distribution along their horizontal central line. In Fig.~\ref{fig:sops} for each point on the sphere, the SOP of the vectorial vortices obtained for $\ell = \pm$1, $\pm$2 and the same values of $(\mathrm{E}_{0x},\mathrm{E}_{0y})$ are also shown. 

Consistent with the above polarisation description, the functions $\Phi_x$ and $\Phi_y$ for vectorial SOPs can also be re-written in terms of $\mathrm{E}_{0x}$ and $\mathrm{E}_{0y}$:
\begin{subequations}
    \begin{gather}
        \Phi_x(\phi) = \frac{1}{\sqrt{2}}\qty\Big[\mathrm{E}_{0x}\,\cos(\ell\phi)-i\mathrm{E}_{0y}\sin(\ell\phi)] \,,\\
        \Phi_y(\phi) = \frac{1}{\sqrt{2}}\qty\Big[\mathrm{E}_{0x}\sin(\ell\phi)+i\mathrm{E}_{0y}\cos(\ell\phi)]  \,.
    \end{gather}
\end{subequations}

\subsection{Electric field evaluation}

Adopting the same piecewise definition of the electric field as in Eq.~\eqref{eq:piecewise_field}, the electric field distribution can be calculated as 
\begin{subequations}
\begin{align}
\va{E}_\mathrm{i}(\va{r}) &= -\frac{ikfe^{-ikf}}{2\pi}\nonumber\\
&\times\sum\limits_{\alpha=s,p}\int\limits_{0}^{\theta_{\mathrm{max}}}\int\limits_{0}^{2\pi}\va{E}_\infty^{(\alpha)}(\theta,\phi)\, e^{i\va{k}^+_\alpha\vdot\va{r}}\sin\theta\,\dd\phi\dd\theta \,,\\
\va{E}_{\mathrm{r}j}(\va{r}) & = -\frac{ikfe^{-ikf}}{2\pi}\nonumber\\
&\times\sum\limits_{\alpha=s,p}\int\limits_{0}^{\theta_{\mathrm{max}}}\int\limits_{0}^{2\pi}r_\mathrm{j}^{(\alpha)}(\theta)\,\va{E}_\infty^{(\alpha)}(\theta,\phi)\, e^{i\va{k}^-_\alpha\vdot\va{r}}\sin\theta\,\dd\phi\dd\theta\,,\\
\va{E}_{\mathrm{t}j}(\va{r}) &=-\frac{ikfe^{-ikf}}{2\pi}\nonumber\\
&\times\sum\limits_{\alpha=s,p}\int\limits_{0}^{\theta_{\mathrm{max}}}\int\limits_{0}^{2\pi}t_\mathrm{j}^{(\alpha)}(\theta)\,\va{E}_\infty^{(\alpha)}(\theta,\phi)\, e^{i\va{k}^+_\alpha\vdot\va{r}}\sin\theta\,\dd\phi\dd\theta\,,
\end{align}
\end{subequations}
where the subscript $j$ labels the region of space and $\alpha$ labels the $s$ and $p$ components of each vector. The term $\va{E}_\infty$ is obtained substituting Eq.~\eqref{eq:Ei_gen} into Eq.~\eqref{eq:focus_field}b, once the functions $\Theta$ and $\Phi$ in Eq.~\eqref{eq:Ei_gen} have been determined for the chosen beam type. Upon substitution of $\va{k}^\pm$ and of the Fresnel coefficients, expressions depending on polarisation (i.e., $s$ or $p$) and layer (i.e., being it isotropic or uniaxial) are obtained. Calculations, limited to the analytical integration over $\phi$, have been written in Wolfram Mathematica, while the final steps have been carried out on MATLAB. Below, the procedure followed for each of the fields appearing in the piece-wise definition of Eq.(\ref{eq:piecewise_field}) is described, using as a mock example the $x$ component of the field $\va{E}_{\mathrm{t}1}$, for an incident field given by a generic VVB, whose SOP is given by $(\mathrm{E}_{0x},\,\mathrm{E}_{0y})$ (assuming it to be normalised to one). In the following,  the electric field taken as an example will be simply labelled as $\va{E}$ for simplicity of notations.

 \begin{table*}[t]
 \setlength\extrarowheight{4pt}
 \captionsetup{type = table}
 \caption{Functions $\zeta^{(s,p)}_{i,\pm}(\theta)$ and $\eta^{(s,p)}_i(\theta)$ in the integrals Eq.~\ref{eq:rw_aniso_integrals} calculated in the medium $i$, for either $s$ or $p$ polarisation and propagation along $\pm \vu{z}$ direction. Labels of the Fresnel coefficients have the same meaning as previously introduced.}
\begin{tabular}{c|cc|cc}
\hline
\multicolumn{5}{c}{\cellcolor{mygray}$\mathbf{\zeta^{(s,p)}_{i,\pm}(\theta)}$ } \\[3pt] \hline
Medium 1              & \multicolumn{4}{c}{$\zeta^{(s,p)}_{1,\pm}(\theta) = \zeta_{1,\pm}(\theta) =\exp\qty\Big(\pm i\,k_0 z \cos\theta \sqrt{\varepsilon_1\mu_1})$}\\[3pt]
\arrayrulecolor{mygray}\hline
 Medium 2             & \multicolumn{4}{c}{$\zeta^s_{2,\pm}(\theta) =\exp\qty\Bigg(\pm i\, k_0 z \sqrt{\frac{\varepsilon_1\mu_1\qty\big(\cos2\theta-1)}{2}+\varepsilon_{2x}\mu_{2x}}),\quad \zeta^p_{2,\pm}(\theta) =\exp\qty\Bigg(\pm i\, k_0 z \sqrt{\varepsilon_1\mu_1\qty\big(\cos2\theta-1)\frac{\varepsilon_{2x}}{2\varepsilon_{2z}}+\varepsilon_{2x}\mu_{2x}})$}\\[3pt]\hline
 Medium 3              & \multicolumn{4}{c}{$\zeta^{(s,p)}_{3,\pm}(\theta) = \zeta_{3,\pm}(\theta)=\exp\qty\Bigg(\pm i\, k_0 z \sqrt{\frac{\varepsilon_1\mu_1\qty\big(\cos2\theta-1)}{2}+\varepsilon_{3}\mu_{3}})$}\\[3pt] 
 \arrayrulecolor{black}\hline
 \multicolumn{5}{c}{\cellcolor{mygray}$\mathbf{\boldsymbol{\eta^{(s,p)}_i(\theta)}}$ } \\[3pt]
 \hline
\multicolumn{1}{c}{}     & \multicolumn{2}{c}{\textbf{Reflection}}                 & \multicolumn{2}{c}{\textbf{Transmission}}      \\[3pt] \hline\arrayrulecolor{mygray}
Medium 1                        & $\quad r_{\mathrm{I}}^s\quad$          & $r_{\mathrm{I}}^p\,\cos\theta$          & $\quad$-$\quad$           & -             \\ [3pt]\hline
Medium 2                        & $r_{\mathrm{II}}^s$         & $r_{\mathrm{I}}^p\,\sqrt{\qty\Big[\varepsilon_1\mu_1\qty\big(\cos2\theta -1)]\varepsilon_{2x}}$                & $t_{\mathrm{I}}^s$         & $t_{\mathrm{I}}^p\,\sqrt{\qty\Big[\varepsilon_1\mu_1\qty\big(\cos2\theta -1)]\varepsilon_{2x}} $         \\ [3pt]\hline
Medium 3    & -            & -                      & $t_{\mathrm{II}}^s$         &  $t_{\mathrm{II}}^p\, \sqrt{\varepsilon_1\mu_1\qty\big(\cos2\theta-1)+2\varepsilon_3\mu_3}$          \\[3pt]\arrayrulecolor{black}\hline
\end{tabular}\label{tab:int_terms}
\end{table*}   
    
In the first step, the field on the lens plane ($\va{E}_\infty$) is divided into its $s$ and $p$ components (Eq.~(\ref{eq:focus_field}b)) and the projections on the basis vectors are calculated:
    \begin{widetext}
    \begin{subequations}
        \begin{align}
            \mathrm{E}_\infty^s(\theta,\phi) & = \va{E}_\mathrm{L}(\theta,\phi)\vdot\vu{n}_\varphi(\theta,\phi) = \bigg(\mathrm{E}_{0y}\cos[(1-\ell)\phi]-\mathrm{E}_{0x}\sin[(1-\ell)\phi]\bigg)\Theta(\theta)\,,\\
            \mathrm{E}_\infty^p(\theta,\phi) & = \va{E}_\mathrm{L}(\theta,\phi)\vdot\vu{n}_\rho(\theta,\phi) =\bigg(\mathrm{E}_{0x}\cos[(1-\ell)\phi]+\mathrm{E}_{0y}\sin[(1-\ell)\phi]\bigg)\Theta(\theta)\,.
        \end{align}
    \end{subequations}
    \end{widetext}
    They are then multiplied by the unit vectors given in Eq.~\eqref{eq:uniax_basis}, according to the studied case, in order to obtain the full vector $\va{E}_\infty^{s,p}$. After that, for each component of the angular spectrum, the factors solely depending on $\theta$ are collected together, yielding:
    \begin{widetext}
    \begin{subequations}
        \begin{gather}
            \mathrm{E}_{\infty,x}^s(\theta,\phi) = \mathrm{C_0}\, \sqrt{\cos\theta}\sin\theta\,\Theta(\theta)\,\vb{E}_\infty^s(\theta,\phi)\sin\phi\,\Xi(\theta,\phi;\varphi)\,,\\
            \mathrm{E}_{\infty,x}^p(\theta,\phi) = \mathrm{C_0}\, \sqrt{\cos\theta}\sin\theta\,\Theta(\theta)\,\vb{E}_\infty^p \sqrt{\frac{\mu_1\varepsilon_1(\cos2\theta-1)}{2\mu_2\varepsilon_{2z}}+1}\cos\phi\,\Xi(\theta,\phi;\varphi)\,,
        \end{gather}
    \end{subequations}
    \end{widetext}
    with $\Xi(\theta,\phi;\varphi) = \exp\bigl[i\sqrt{\varepsilon_1\mu_1}\rho k_0 \cos(\phi-\varphi)\sin\theta\bigr]$, $C_0 = \mathrm{C}\sqrt{\mu_1\varepsilon_1}$.
    In the above expressions, the term $\sqrt{\cos\theta}$ comes from the intensity  law, while the term $\sin\theta$ (which is in addition to the one contained in $\Theta$) comes from the Jacobian of the integration variables transformation. In the next step, the integration over $\phi$ is carried out applying Eq.~\eqref{eq:besseltrig} and upon simplification of the trigonometric functions of $\phi$, when necessary. 
    With Wolfram Mathematica, it is indeed possible to integrate Bessel functions of the first kind automatically and by means of Eq.(\ref{eq:besseltrig}), but this is no longer possible if the order of the Bessel function is kept arbitrary. This would imply the need of specifying the desired order of the Laguerre polynomial before completing the first integration. For this reason, the code also includes a list of substitutions used to apply Eq.(\ref{eq:besseltrig}) to a wide set of combinations of trigonometric functions, returning the corresponding combination of Bessel functions. The list has been built using all the possible combinations found in the calculations, to ensure the possibility to write a general solution for any arbitrary beam LG$_{\ell p}$.   
    Upon completion of this step, the field spectral components become:
    \begin{widetext}
    \begin{subequations}
        \begin{align}
            \mathrm{E}^s_{\infty,x}(\theta;\varphi) &=\pi  \mathrm{C} k_0  i^{-\ell } \sin (\theta ) \sqrt{\cos (\theta )} \Theta(\theta) \sqrt{\mu_1 \varepsilon_1} \nonumber\\
            &\times\biggl[\biggl(\mathrm{E}_{0x} \cos\Bigl[\varphi(\ell -2)\Bigr]-\mathrm{E}_{0y} \sin \Bigl[\varphi(\ell-2)\Bigr]\biggr) J_{2-\ell }\Bigl(\sqrt{\varepsilon_1 \mu_1} \rho  k_0 \sin\theta\Bigr)\nonumber\\
            &+i^{2 \ell } \biggl(\mathrm{E}_{0x} \cos \ell\varphi-\mathrm{E}_{0y} \sin \ell\varphi\biggr) J_{\ell }\Bigl(\sqrt{\varepsilon_1 \mu_1} \rho  k_0 \sin\theta\Bigr)\biggr]\,,\\
            \mathrm{E}^p_{\infty,x}(\theta;\varphi) &=\pi  \mathrm{C} k_0 i^{-\ell } \sin (\theta ) \sqrt{\cos (\theta )} \Theta(\theta) \sqrt{\mu_1 \varepsilon_1} \sqrt{\frac{(\cos (2 \theta )-1) (\mu_1 \varepsilon_1)}{2 \mu_2 \varepsilon_{2z}}+1} \nonumber\\
            &\times\biggl[\biggl(\mathrm{E}_{0y} \sin \Bigl[\varphi(\ell-2)\biggr]-\mathrm{E}_{0x} \cos \Bigl[\varphi(\ell-2)\Bigr]\Bigr) J_{2-\ell }\Bigl(\sqrt{\varepsilon_1 \mu_1} \rho  k_0 \sin\theta\Bigr)\nonumber\\
            &+i^{2 \ell } \Bigl(\mathrm{E}_{0x} \cos \ell\varphi-\mathrm{E}_{0y} \sin \ell\varphi\Bigr) J_{\ell }\Bigl(\sqrt{\varepsilon_1 \mu_1} \rho  k_0 \sin\theta\Bigr)\biggr]\,.            
        \end{align}
    \end{subequations}
    \end{widetext}
    
    The components are then re-arranged collecting the Bessel functions involved. The resulting expressions are used in five integrands with which all the field components can be reconstructed:
    \begin{widetext}
    \begin{subequations}\label{eq:rw_aniso_integrals}
    \begin{align}
      \mathcal{I}_{1} &=\int_0^{\theta_\mathrm{max}}\,J_\ell\Bigl(k_0\sqrt{\varepsilon_1\mu_1}\rho\sin\theta\Bigr)\,f_w(\theta) \sqrt{\cos\theta}\sin\theta\,\qty\big(\cos\theta+1)\,\Theta(\theta)\,\zeta^{(s,p)}_{i,\pm}(\theta)\,\eta^{(s,p)}_i(\theta)\,\dd\theta \\  
      \mathcal{I}_{2} &=\int_0^{\theta_\mathrm{max}}\,J_{2-\ell}\Bigl(k_0\sqrt{\varepsilon_1\mu_1}\rho\sin\theta\Bigr)\,f_w(\theta) \sqrt{\cos\theta}\sin\theta\,\qty\big(\cos\theta-1)\,\Theta(\theta)\,\zeta^{(s,p)}_{i,\pm}(\theta)\,\eta^{(s,p)}_i(\theta)\,\dd\theta \\  
      \mathcal{I}_{3} 
     &=\int_0^{\theta_\mathrm{max}}\,J_{2+\ell}\Bigl(k_0\sqrt{\varepsilon_1\mu_1}\rho\sin\theta\Bigr)\,f_w(\theta) \sqrt{\cos\theta}\sin\theta\,\qty\big(\cos\theta-1)\,\Theta(\theta)\,\zeta^{(s,p)}_{i,\pm}(\theta)\,\eta^{(s,p)}_i(\theta)\,\dd\theta \\
      \mathcal{I}_{4} &=\int_0^{\theta_\mathrm{max}}\,J_{1-\ell}\Bigl(k_0\sqrt{\varepsilon_1\mu_1}\rho\sin\theta\Bigr)\,f_w(\theta) \sqrt{\cos\theta}\sin^2\theta\,\Theta(\theta)\,\zeta^p_{i,\pm}(\theta)\,\eta^p_i(\theta)\,\dd\theta \\
     \mathcal{I}_{5} &=\int_0^{\theta_\mathrm{max}}\,J_{1+\ell}\Bigl(k_0\sqrt{\varepsilon_1\mu_1}\rho\sin\theta\Bigr)\,f_w(\theta) \sqrt{\cos\theta}\sin^2\theta\,\Theta(\theta)\,\zeta^p_{i,\pm}(\theta)\,\eta^p_i(\theta)\,\dd\theta\,.
    \end{align}      
    \end{subequations}
    \end{widetext}
    The explicit form of $\Theta$ depends on the beam type. 
    The functions $\zeta(\theta)$ and $\eta(\theta)$ represent quantities whose explicit expression depends on the medium (labelled by $i$), polarisation ($s$ or $p$) and propagation direction ($\pm\vu{z}$) they are calculated for (Table~\ref{tab:int_terms}). The former ($\zeta$) is the propagation factor and the latter ($\eta$) is a general representation of the Fresnel coefficients needed in each case. It is worth noting that in the case of the integrals $\mathcal{I}_{4}$ and $\mathcal{I}_{5}$, only the contribution of the $p$ component of the field is important. Finally, the numerical integration is performed in MATLAB.
 \begin{figure*}[!t]
    \centering
    \includegraphics[scale =1]{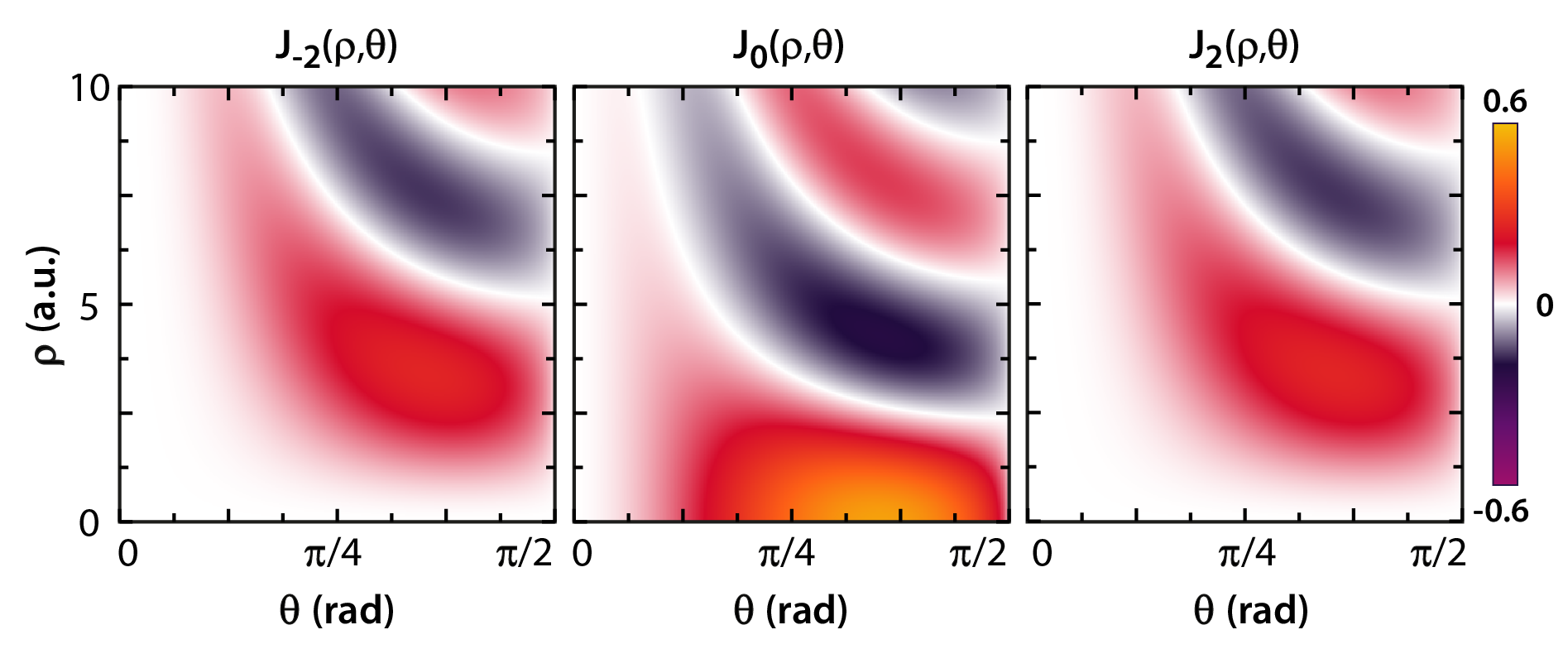}
    \caption{Maps of the integrands $\xi_{\ell\pm1}$ in Eqs.~\eqref{eq:rw_aniso_integrals} plotted for $\ell = -2,0,2$. 
    The colour scale is the same for the three maps.}
    \label{fig:integrands}
\end{figure*}
 Similar procedure can be applied to obtain the magnetic field, retrieving its angular spectrum from the electric field one.

 \begin{table*}
\setlength\extrarowheight{4pt}
\captionsetup{type = table}
\caption{Expression of the electric field $\va{E}_{t1}$ as an example in the description of the methodology. The vector components are given for both cases of a scalar (OVB) and vectorial (VVB) vortex beams of generic polarisation state $(\mathrm{E}_{0x},\mathrm{E}_{0y})$ and topological charge $\ell$. Some terms appearing in the field components have been gathered in extra factors, whose definition is given in the bottom part of the table. All the integrals are labelled with a superscript showing the medium they are calculated in and the polarisation component they refer to. In each integral, the Fresnel coefficients in $\eta$ represent transmission coefficients, according to Table~\ref{tab:int_terms}.}
\begin{tabular}{cl|l}
\hline
\multicolumn{3}{c}{\cellcolor{mygray}{ \textbf{Electric field in medium 2, propagating along positive} $\mathbf{z}$}} \\ [4pt]\hline\arrayrulecolor{mygray}
\multicolumn{1}{c|}{}                       &                      & $ \mathrm{C_0}\,\Big[\mathrm{E}_{0x}\qty\big(f_{\varphi}^{10}-if_{\varphi}^{20})\qty\Big(\sqrt{2}\mathcal{I}_1^{1p} + 2\sqrt{\varepsilon_{2x}\varepsilon_{2z}\mu_2}\mathcal{I}_1^{1s}) + \frac{i^{-2\ell}}{2}(i\mathrm{E}_{0x} +\mathrm{E}_{0y}) \qty\big(i f_{\varphi}^{12}+f_{\varphi}^{22})\qty\Big(\sqrt{2}\mathcal{I}_2^{1p} - 2\sqrt{\varepsilon_{2x}\varepsilon_{2z}\mu_2}\mathcal{I}_2^{1s})+$  \\[5pt]
\multicolumn{1}{c|}{}                       & \multirow{-2}{*}{$\, x\,$}  & $-\frac{1}{2}(\mathrm{E}_{0x} +i\mathrm{E}_{0y})\qty\big(f_{\varphi}^{13}-if_{\varphi}^{23})\qty\Big(\sqrt{2}\mathcal{I}_3^{1p} - 2\sqrt{\varepsilon_{2x}\varepsilon_{2z}\mu_2}\mathcal{I}_3^{1s})\Big]$  \\[5pt]\cline{3-3} 
\multicolumn{1}{c|}{}                       &                      &  $ \mathrm{C_0}\,\Big[\mathrm{E}_{0y}\qty\big(f_{\varphi}^{10}-if_{\varphi}^{20})\qty\Big(\sqrt{2}\mathcal{I}_1^{1p} + 2\sqrt{\varepsilon_{2x}\varepsilon_{2z}\mu_2}\mathcal{I}_1^{1s}) + \frac{i^{-2\ell}}{2}(i\mathrm{E}_{0x} +\mathrm{E}_{0y}) \qty\big(f_{\varphi}^{12}-if_{\varphi}^{22})\qty\Big(\sqrt{2}\mathcal{I}_2^{1p} - 2\sqrt{\varepsilon_{2x}\varepsilon_{2z}\mu_2}\mathcal{I}_2^{1s})+$  \\ [5pt]
\multicolumn{1}{c|}{}                       & \multirow{-2}{*}{$\, y\,$}  &   $-\frac{1}{2}(i\mathrm{E}_{0x} -\mathrm{E}_{0y})\qty\big(f_{\varphi}^{13}-if_{\varphi}^{23})\qty\Big(\sqrt{2}\mathcal{I}_3^{1p} - 2\sqrt{\varepsilon_{2x}\varepsilon_{2z}\mu_2}\mathcal{I}_3^{1s})\Big]$   \\ [5pt]\cline{3-3} 
\multicolumn{1}{c|}{\multirow{-6}{*}{OVB}}  & $\, z\,$                    &  $\mathrm{C_0}\,\frac{2\,i^{-2\ell}\varepsilon_{2x}\sqrt{\varepsilon_1\mu_1}}{\sqrt{\varepsilon_{2z}}}\qty\big(f_{\varphi}^{10}-if_{\varphi}^{20})\qty\Big[\qty\big(\mathrm{E}_{0x}-i\mathrm{E}_{0y})\qty\big(-if_{\varphi}^{01}+f_{\varphi}^{02})\mathcal{I}_4^{1p}+i^{2\ell}\qty\big(-i\mathrm{E}_{0x}+i\mathrm{E}_{0y})\qty\big(f_{\varphi}^{01}-if_{\varphi}^{02})\mathcal{I}_5^{1p}]$ \\ [5pt]\arrayrulecolor{black}\cline{1-3} \arrayrulecolor{mygray}
\multicolumn{1}{l|}{}                       & $\, x\,$                    &  $ \mathrm{C_0}\qty\Big[\,\qty\big(\mathrm{E}_{0x} f_{\varphi}^{10}-\mathrm{E}_{0y}f_{\varphi}^{20})\qty\Big(\sqrt{2}\mathcal{I}_1^{1p} + 2\sqrt{\varepsilon_{2x}\varepsilon_{2z}\mu_2}\mathcal{I}_1^{1s})-i^{-2\ell}\qty\big(\mathrm{E}_{0x}f_{\varphi}^{12}-\mathrm{E}_{0y}f_{\varphi}^{22})\qty\Big(\sqrt{2}\mathcal{I}_2^{1p}-2\sqrt{\varepsilon_{2x}\varepsilon_{2z}\mu_2}\mathcal{I}_1^{1s})]$ \\[5pt] \cline{3-3} 
\multicolumn{1}{l|}{}                       & $\, y\,$                    &   $ \mathrm{C_0}\,\qty\Big[\qty\big(\mathrm{E}_{0x} f_{\varphi}^{10}+\mathrm{E}_{0y}f_{\varphi}^{20})\qty\Big(\sqrt{2}\mathcal{I}_1^{1p} + 2\sqrt{\varepsilon_{2x}\varepsilon_{2z}\mu_2}\mathcal{I}_1^{1s})+i^{-2\ell}\qty\big(\mathrm{E}_{0x}f_{\varphi}^{12}+\mathrm{E}_{0y}f_{\varphi}^{22})\qty\Big(\sqrt{2}\mathcal{I}_2^{1p}-2\sqrt{\varepsilon_{2x}\varepsilon_{2z}\mu_2}\mathcal{I}_1^{1s})]$\\[5pt] \cline{3-3} 
\multicolumn{1}{l|}{\multirow{-3}{*}{VVB}}  & $\, z\,$                    &  $-4\mathrm{C_0}\,i^{1-2\ell}\sqrt{\varepsilon_1\varepsilon_{2x}\mu_1}{\sqrt{\varepsilon_1\varepsilon_{2z}\mu_1}} \qty\big(\mathrm{E}_{0x} f_{\varphi}^{11}+\mathrm{E}_{0y}f_{\varphi}^{21}) \mathcal{I}_4^{1p}$ \\[5pt] \cline{1-1} \cline{2-3} 
 \arrayrulecolor{black}\hline
\multicolumn{3}{c}{\cellcolor{mygray}{ \textbf{Factors appearing in the expressions above}}} \\ [4pt]\hline\arrayrulecolor{mygray}
\multicolumn{3}{c}{$\mathrm{C_0} = i^\ell\pi\mathrm{C}k_0\frac{\sqrt{\varepsilon_1\mu_1}}{2\sqrt{\varepsilon_{2x}\varepsilon_{2z}\mu_2}}$}\\[5pt]\hline\arrayrulecolor{mygray}
\multicolumn{3}{c}{$f_{\varphi}^{01} = \cos\varphi\qquad f_{\varphi}^{02} = \sin\varphi$}\\[5pt]\hline
\multicolumn{3}{c}{$f_{\varphi}^{10} = \cos\ell\varphi\qquad f_{\varphi}^{11} = \cos\qty\Big[\qty\big(1-\ell)\varphi]\qquad f_{\varphi}^{12} = \cos\qty\Big[\qty\big(2-\ell)\varphi]\qquad f_{\varphi}^{13} = \cos\qty\Big[\qty\big(2+\ell)\varphi]$}\\[5pt]\hline
\multicolumn{3}{c}{$f_{\varphi}^{20} = \sin\ell\varphi\qquad f_{\varphi}^{21} = \sin\qty\Big[\qty\big(1-\ell)\varphi]\qquad f_{\varphi}^{22} = \sin\qty\Big[\qty\big(\ell-2)\varphi]\qquad f_{\varphi}^{23} = \sin\qty\Big[\qty\big(2+\ell)\varphi]$}\\[5pt] \arrayrulecolor{black}\hline
\end{tabular}
\label{tab:rwt_ex_field}
\end{table*}

To provide a practical example, Table~\ref{tab:rwt_ex_field} presents the expressions obtained for the electric field $\va{E}_{t1}$ in both the cases of OVs and VVBs, where the dependence on the coordinates $(\rho, \varphi,z)$ have been omitted for simplified notations.

It is worth noting that, when calculating the field for a VVB, the three $\theta$-integrals ($\mathcal{I}_1$, $\mathcal{I}_2$, and $\mathcal{I}_4$) need to be evaluated, while the five integrals are needed for a scalar vortex. This is related to the presence of the $\phi$-dependent phase factor of the LG modes, which is lost in the superposition of orthogonal circular vortices with opposite topological charge, needed for VVBs.
 
\subsection{Longitudinal field example}

This semi-analytical approach introduces an extension of the Richards-Wolf theory of vectorial diffraction to an anisotropic medium. The advantage of a semi-analytical model, compared to fully numerical studies, is the possibility to handle close expressions for the fields and retrieve the related fundamental information from the functions describing their components. For example, it is interesting to look at the longitudinal ($z$-) component generated by the focusing. In both cases of OVs and VVBs, it solely depends on the $p$ polarisation component of the angular spectrum, since it only contains the integrals $\mathcal{I}_4$ and $\mathcal{I}_5$. While the spatial distribution of the the longitudinal field depends on the results of the integration and on all the factors contributing to it, it is possible to understand the general trends directly from the integrands. In particular, $\mathcal{I}_4$ and $\mathcal{I}_5$ contain the terms $\xi_{1\pm\ell}(\rho,\theta) = J_{1\pm\ell}\Bigl(k_0\sqrt{\mu_1\varepsilon_1}\rho\sin\theta\Bigr)\sqrt{\cos\theta}\sin^2\theta$, which are calculated for each value of $\rho$ over all the range of values of the integration variable $\theta$, so that the field at any point in space, depends on the superposition of all the plane waves entering the field of view of the objective. 

The integrals $\mathcal{I}_4$ and $\mathcal{I}_5$ show a case for $\ell = \pm1$, which leads to the appearance of the Bessel function $J_0$. This is in fact the only function of the set being nonzero-valued when its argument is zero, which implies that only the terms containing this function will have a non-zero integral in the spatial points corresponding to $\rho=0$. Comparing the maps of $\xi_\ell(\rho,\theta)$ for $\ell= 0,\pm1$ (Fig.~\ref{fig:integrands}), it is evident how $J_{0}$ is the only case where for small values of $\rho$ the integral is nonzero. Looking at the same maps for higher values of $\rho$, it is expected that the value of these integrals will show regions of different signs for the increasing distance from the origin.

Considering the case of OVs, the $z$ component of the focused field contains the Bessel function $J_0$ for $\ell = -1$ and this appears together with the function $J_2$, making more detailed predictions quite cumbersome at this stage, given all the complicated dependencies appearing in the integrals and the general dependence on the SOP. On the other hand, when assuming $\ell=1$ for a VVB on a higher-order Poincaré sphere, the $z$ component only contains the Bessel function $J_0$ and two particular cases can be highlighted. The $z$ component of the VVB field when the dependence on the coordinate $\varphi$ is made explicit is: 
\begin{widetext}
\begin{subequations}
\begin{align}
    \mathrm{E}^{(\mathrm{VVB})}_z(\rho,\varphi,z) 
    &= -4i\varepsilon_{2z}\sqrt{\mu_1\varepsilon_1}\mathrm{C_0}\,J_{1-\ell}\Bigl(k_0\sqrt{\mu_1\varepsilon_1}\rho\sin\theta\Bigr)\sqrt{\cos\theta}\sin^2\theta\nonumber\,,\\
    &\times\biggl(\mathrm{E}_{0x}\cos\Bigl[(1-\ell)\varphi\Bigr]+ \mathrm{E}_{0y}\sin\Bigl[(1-\ell)\varphi\Bigr]\biggr)\\  \biggl\{\ell=1\implies\biggr\}&=-4i\varepsilon_{2z}\sqrt{\mu_1\varepsilon_1}\mathrm{C_0}\,J_0\Bigl(k_0\sqrt{\mu_1\varepsilon_1}\rho\sin\theta\Bigr)\sqrt{\cos\theta}\sin^2\theta\,\mathrm{E}_{0x}\,,
\end{align}
\end{subequations}
\end{widetext}
so that the only contribution to the longitudinal field can come from the component of the beam on the state $\ket{\mathrm{H},1}$ ($(\mathrm{E}_{0x},0)$, in this case). This state of polarisation represents a radial beam, which is known for its nonzero longitudinal field, whose intensity can be strongly increased by tight focusing~\cite{quabis2000focusing,scully1991simple,dorn2003sharper}.
On the other hand, if the initial SOP is given by $(0, \mathrm{E}_{0y})$, which corresponds to an azimuthal beam, the longitudinal field will be zero in every point of the real space.
\begin{figure*}[!t]
    \centering
    \includegraphics[scale = 1]{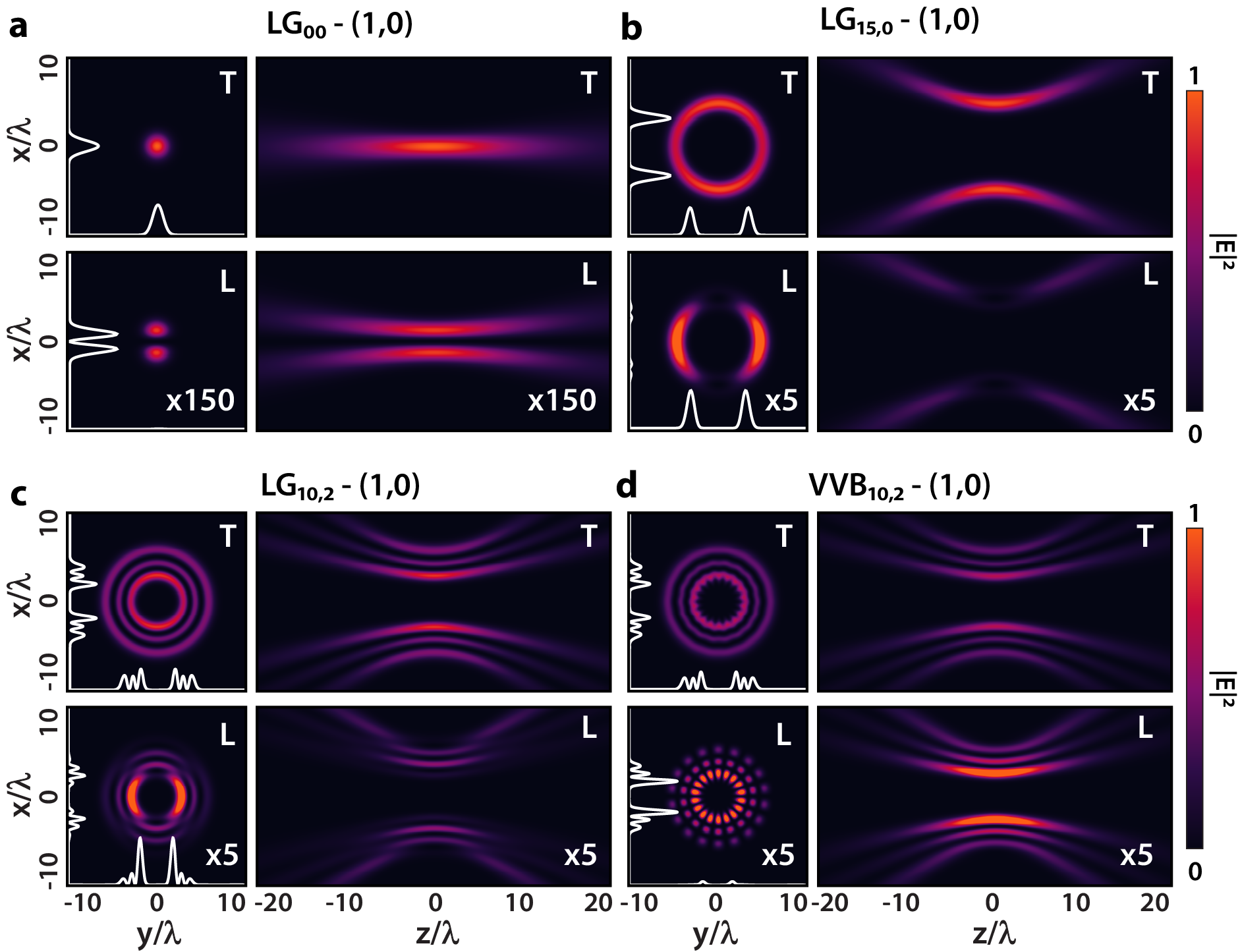}
    \caption{Vortex beam focusing in free space with NA = $0.9$: (a) Gaussian beam, (b) scalar vortex beam LG$_{15,0}$,  (c) scalar vortex beam LG$_{10,2}$, (d) vectorial vortex beam LG$_{\pm10,2}$. Colour maps represent the intensity of the (T) transverse and (L) longitudinal electric field components, calculated in both (left) transverse ($x-y$) at $z=0$ and (right) longitudinal ($x-z$) at $y=0$ planes. The line plots are the cross-sections of the corresponding maps along the $x$ or $y$) axes. The SOP of each type of beam is shown in brackets, according to the basis used for scalar and vectorial beams. }
    \label{fig:fs}
\end{figure*}

\section{Application examples}
The approach described in the previous section provides the possibility to simulate the focusing of different types of optical vortices and their propagation through a three-layered medium. The required inputs are: (i) $\mathrm{E}_{0x}$, $\mathrm{E}_{0y}$, $ell$, $p$, $w_0$ and $\lambda$ in order to completely characterise the beam; (ii) the permittivities of each layer, to describe its optical behaviour; and (iii) the numerical aperture and the location of the geometrical focus of the lens with respect to the interfaces, to characterise the focusing features of the system.        
We consider several archetypal cases in order to demonstrate the versatility of the developed approach:
a standard Gaussian beam, obtained as a Laguerre-Gauss beam with both indices set to 0 (LG$_{00}$); a high-order scalar vortex (LG$_{15,0}$); a scalar vortex with a nonzero radial index (LG$_{10,2}$); and a vectorial vortex whose constituent vortices are given by LG$_{\pm10,2}$ beams, referred to as the VVB$_{10,2}$ (Fig.~\ref{fig:fs}). The wavelength of choice is $\lambda $ = 650 nm and the beams are focused with an objective of numerical aperture of 0.9 in all cases. All the scalar beams have been simulated assuming horizontal polarisation (i.e., the electric field parallel to the $\vu{x}$ direction), hence only one of the initial electric field components is nonzero ($\mathrm{E}_{0y} = 0$). The vectorial vortex is calculated with the same prescription, so that its polarisation state can be thought of as a higher-order equivalent of the horizontal scalar polarisation: scalar and vectorial are located on the same point of the Poincaré sphere, but obtained for different values of topological charge. Any other state on the Poincaré sphere is obtained by a careful selection of $\mathrm{E}_{0x}, \mathrm{E}_{0y}$ and $\ell$ (cf. Fig.~\ref{fig:sops}). 

\paragraph{\textbf{Free space.}} The simplest case to model is the propagation of the focused beam in free space, where all the permittivities are set equal to one (Fig.~\ref{fig:fs}). The size of the spatial mesh used in the calculations has been expanded to fully appreciate the spatial variations of the transverse and longitudinal field components, over a macroscopic distance. Given the strong focusing regime, even a simple Gaussian beam develops a nonzero longitudinal field, which shows a two-lobed shape, aligned to the polarisation direction.  Although nonzero, the longitudinal field intensity is considerably smaller than the transverse one: the maximum value of the former is approximately $0.6\%$ of the latter. If a nonzero topological charge is introduced, an immediate rise in the longitudinal field strength is observed. Its maximum becomes in fact the $26\%$ of the transverse field for the scalar LG$_{15,0}$ vortex, reaching a $35\%$ for LG$_{10,2}$. Common to all the scalar vortices, not limited to the three presented here, the symmetry of the longitudinal field intensity distribution is affected by a focusing-induced astigmatism, which is manifested in a prolate shape of the beam. This is different from a vectorial beam, where the interference between the co-propagating vortices of opposite topological charges results in a cylindrically symmetric shape. Together with a higher degree of symmetry, the intensity of the longitudinal field relative to its transverse counterpart is also increased by the vectorial nature of this types of beam: the longitudinal field maximum becomes approximately $50\%$ of the transverse one, making these beams appealing for applications where a strong longitudinal field is needed. 
\begin{figure*}[t!]
    \centering
    \includegraphics[scale = 1]{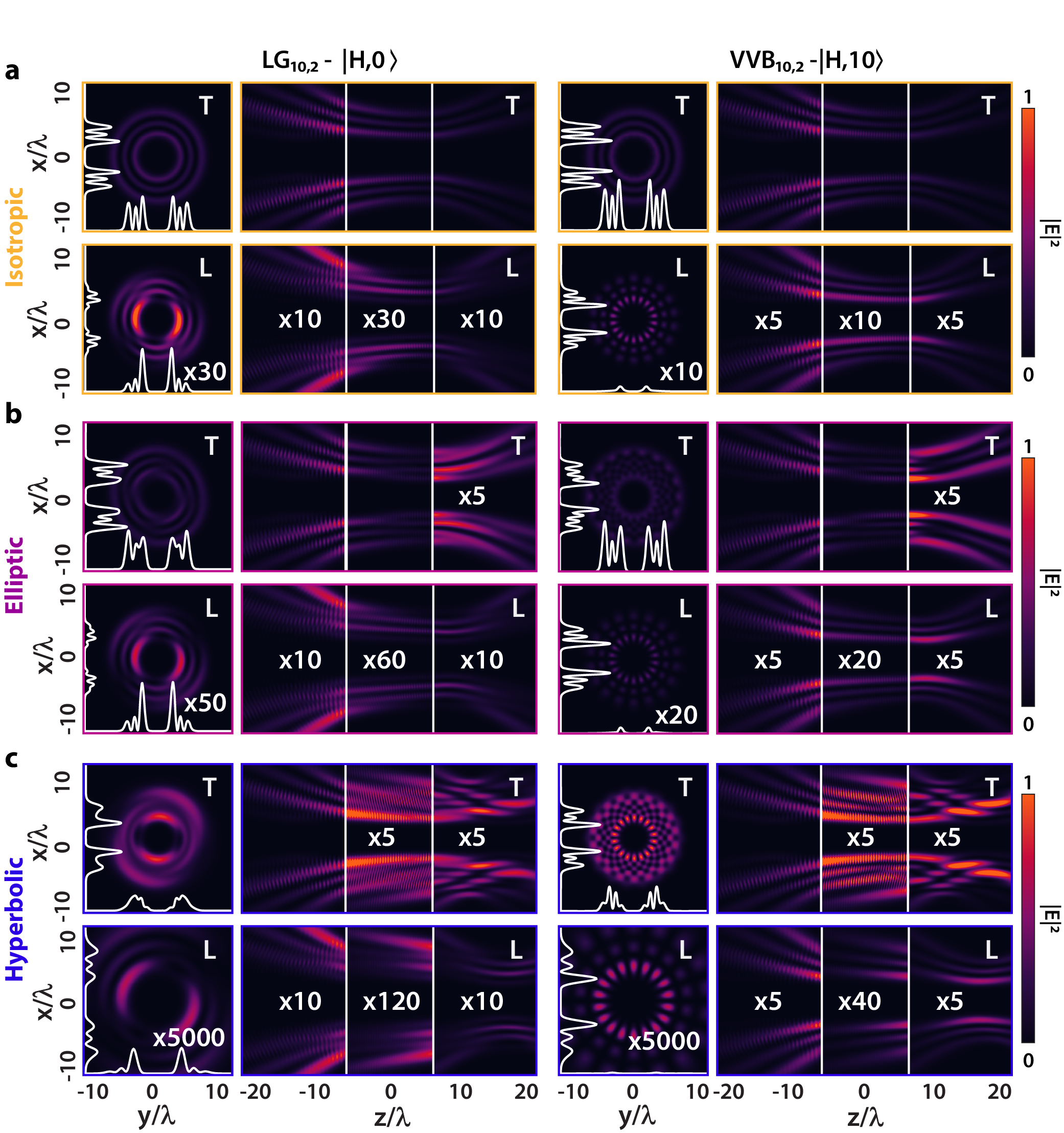}
    \caption{Propagation of (a) LG$_{10,2}$ and (b) VVB$_{10,2}$ beams focused with NA = 0.9 in a three-layered system. The central layer can be a (a, yellow frames) isotropic ($\varepsilon_{2x} = \varepsilon_{2z} = 2.25 + i 5\times10^{-2} $), (b, purple frames) elliptic ($\varepsilon_{2x} = 2.25 + i 5\times10^{-2}$, $\varepsilon_{2z} = 1.9 + i 5\times10^{-2}$), (c, blue frames) hyperbolic ($\varepsilon_{2x} = 2.25 + i 5\times10^{-2}$, $\varepsilon_{2z} = -1.9 + i 5\times10^{-2}$) medium, while the first and third layers are chosen as free space ($\varepsilon_1=\varepsilon_3$ = 1). The description of the content of the panels is the same as in Fig.~\ref{fig:fs}. The multiplication factors refer to the panel they are shown in. }
    \label{fig:slab} 
\end{figure*}
\paragraph{\textbf{Isotropic lossy slab.}}
The simulated system can be made slightly more complicated by the introduction of a single interface, which can be obtained using the same values of permittivities for two consecutive layers. For the sake of brevity, from here on we are only reporting the results for a more general case of a double interface and limited to the beams LG$_{10,2}$ and VVB$_{10,2}$. Any of the regions of the system (Fig.~\ref{fig:aniso_slab}) can be isotropic and lossy (complex values of permittivity are supported in the developed approach), so that the first case considered is a lossy dielectric (Fig.~\ref{fig:slab}, yellow frames) with permittivity $\varepsilon_{2x} = \varepsilon_{2z} = 2.25 + i 5\times10^{-2}$. The results are similar to the previous case of free-space propagation with the main difference being a reduction of the overall intensity of the beam along the propagation direction, equally damping both the transverse and longitudinal components of the beam. However, the reflection at the slab boundaries results in the field intensity redistribution across the beam cross-section for both field components. 

\paragraph{\textbf{Uniaxial slab: elliptic dispersion.}}
The central layer can also allow anisotropy, as long as there is a single optical axis and it is aligned with the $\vu{z}$ direction of the reference frame, which coincides with the normal to the interfaces. Depending on the sign of the permittivity components $\Re(\varepsilon_{2x})$ and $\Re(\varepsilon_{2z})$ of Eq.~\eqref{eq:wave_aniso}), this case describes different dispersion regimes. When their product is positive, the dispersion regime is a conventional elliptic, inheriting the name from the shape of the $k$-surface characteristic of this case (Fig.~\ref{fig:slab}, purple frames). The permittivities have been chosen to be $\varepsilon_{2x} = 2.25 + i 5\times10^{-2}$, $\varepsilon_{2z} = 1.9 + i 5\times10^{-2}$. In this case, the transverse field, affected by $\varepsilon_{2x}$, shows the same behaviour as in the previous case of a lossy isotropic dielectric. The differences from the isotropic case are visible in the shapes of the intensity distributions of the beam profiles, augmented by a variation of the standing wave pattern obtained upon multiple reflections inside the anisotropic slab. On the other hand, a smaller real part of $\Re(\varepsilon_2z)$, which introduces anisotropy, produces a stronger longitudinal field inside the slab, as a consequence of the conservation of the longitudinal component of the electric displacement vector. The imaginary parts have been kept the same as in the case of an isotropic dielectric in order to avoid overlapping different effects (different values for $\Im(\varepsilon_{2x})$ and $\Im(\varepsilon_{2z})$ would produce different damping rates for transverse and longitudinal components and interplay between them).

\paragraph{\textbf{Uniaxial slab: hyperbolic dispersion.}}
 If eventually $\Re(\varepsilon_{2x})\Re(\varepsilon_{2z})<0$, the second layer has a hyperbolic dispersion (Fig.~\ref{fig:aniso_slab}, blue frames). The chosen permittivities for this case are: $\varepsilon_{2x} = 2.25 + i 5\times10^{-2}$, $\varepsilon_{2z} = -1.9 + i 5\times10^{-2}$, where again the transverse component matches the value used for an isotropic dielectric and all the imaginary parts are kept the same. The model reliably reproduces the effects of negative refraction, phenomenon known
 to happen in hyperbolic materials \cite{veselago1968reviews,pendry2000negative,smith2003electromagnetic}. The beam is in fact being re-focused to a point outside of the uniaxial slab, and its lateral dimensions are strongly modified inside of it. The modifications of the field intensity distributions are also clearly visible similar to the previous cases but with stronger damping of the longitudinal field and stronger divergence of the transverse field inside the slab, which are connected because of the inter-relation between longitudinal and transverse fields in the beam \cite{aita2023,afanasev2023nondiffractive}. 
 
The described approach is obviously not limited to the case of natural materials and can be applied to structured materials if the effective medium considerations can be used to describe optical properties through effective permittivity. For example, depending on the constituent materials and geometric parameters, plasmonic nanorod-based metamaterials or metal-dielectric multilayered metamaterials may exhibit elliptic or hyperbolic dispersion regimes, or an epsilon-near-zero regime, when a permittivity tensor component approaches zero. The developed approach can be used for simulations of reflection, transmission and absorption spectra, and reveals how the polarisation and intensity distributions of a beam are modified by a particular dispersion regime of the metamaterial~\cite{aita2023}.

\section{Conclusions}
The developed semi-analytical model for the vectorial diffraction theory in anisotropic uniaxial media can be applied to a wide range of situations, including focused optical vortices, both scalar and vectorial, and their propagation through a dielectric slab, whose optical behaviour can be either isotropic or uniaxial, exploring physically interesting dispersion cases, such as hyperbolic and epsilon near-zero regimes. The presented approach allows for a comprehensive investigation of various parameters, including the objective numerical aperture, the material permittivity, the slab thickness, and the beam state of polarisation. Moreover, by exploiting the Laguerre-Gauss basis for the description of the vortices, this approach can addresses crucial aspects of optical wave propagation in complex media in regards of orbital angular momentum physics. 

The possibility to model such a system can assist the exploration of a wide range of applications, such as optical communication, imaging systems, and laser beam shaping, where the propagation of an optical vortex through an anisotropic slab can play a crucial role. The simulations described here are provided as an open-source software package called ``InFOCUS'' (Interaction of FOcused Complex beams with Uniaxial Slabs), available at Ref.\cite{Aita2024}.

In addition to the variety of cases the presented model can be applied to, there are still many potential extensions that could be implemented for future improvements. For instance, extending the model to include the possibility to change the angle of incidence of the incoming beam or, on the material side, including cases described by a more complicated dielectric tensor like, for example, chiral media. 

\section*{Acknowledgemnts.} This work was supported in part by the ERC iCOMM Project (No. 789340) and the ERC Starting Grant No. ERC-2016-STG-714151-PSINFONI. All the data supporting finding of this work are presented in the Results section and are available from the corresponding author upon reasonable request.

\bibliography{References.bib}

\clearpage

\end{document}